# Phase diagram of $La_{5/8-y}Nd_yCa_{3/8}MnO_3$ manganites

J Sacanell[1,4], P Levy[1], A G Leyva[1,2], F Parisi[1,2] and L Ghivelder[3]

[1]Departamento de Física, Centro Atómico Constituyentes, CNEA, Av. Gral. Paz 1499 (1650), San Martín, Prov. de Buenos Aires, Argentina.

[2]Escuela de Ciencia y Tecnología, UNSAM, San Martín, Buenos Aires, Argentina.

[3]Instituto de Fisica, Universidade Federal do Rio de Janeiro, Caixa Postal 68528, Rio de Janeiro, RJ 21941-972, Brazil.

[4]Corresponding author: sacanell@cnea.gov.ar

We report a detailed study of the electric transport and magnetic properties of the $La_{5/8-y}Nd_yCa_{3/8}MnO_3$ manganite system. Substitution of $La^{3+}$ by smaller $Nd^{3+}$ ions, reduces the mean ionic radius of the A – site ion. We have studied samples in the entire range between rich La and rich Nd compounds ($0.1 < y < 0.625$). Results of DC magnetization and resistivity show that doping destabilize the FM character of the pure La compound and triggers the formation of a phase separated state at intermediate doping. We have also found evidence of a dynamical behaviour within the phase separated state. A phase diagram is constructed, summarizing the effect of chemical substitution on the system.

PACS: 75.47.Lx

# 1. Introduction

Manganese based oxides, also known as manganites, represent one of the most widely studied strongly correlated electron system since the discovery of the colossal magnetoresistance (CMR) effect [1]. A substantial amount of work was devoted to study the low temperature state of the manganites characterized by the intrinsic coexistence of two or more phases on a submicrometric scale [2], a phenomenon known as phase separation (PS). This mixture is usually formed by ferromagnetic (FM) conductive and antiferromagnetic (AFM) insulating charge ordered (CO) phases [3,4].

It is known that the substitution of La by Pr in the $La_{5/8-y}Pr_yCa_{3/8}MnO_3$ system destabilizes the FM state in favor of the AFM CO one. This effect is attributed to the increment of the distortions on the Mn – O – Mn bonds induced by doping. The electrical and magnetic [5] and the thermal properties [6] of this system can be described within the PS scenario. In addition, several works have reported the existence of a dynamical behaviour in different manganite samples which display the phase separation phenomena [5,7-11]. It is now believed that chemical and structural disorder are the key factors for the formation of a glass-like state [12,13].

The $La_{5/8-y}Nd_yCa_{3/8}MnO_3$ system is a solid solution of $La_{5/8}Ca_{3/8}MnO_3$ and $Nd_{5/8}Ca_{3/8}MnO_3$, leading to a series of compounds in which the only varied parameter is the mean radius of the A – site ion. Nd ions are slightly smaller than Pr ions, so the effects related to the Mn – O – Mn distortions are expected to be enhanced. In the literature of manganites there are few reports on $La_{5/8-y}Nd_yCa_{3/8}MnO_3$. Rao et al. [14] studied the structural features on the $La_{(2-x)/3}Nd_{x/3}Ca_{1/3}MnO_3$ series in which they suggest the existence of La rich metallic regions coexisting with Nd rich semiconducting regions. A *T-z* phase diagram, mostly based on magnetic measurements, was already obtained for the $(La_{1-z}Nd_z)_{1-x}Ca_xMnO_3$ system (x=0.45), showing a change from ferromagnetic metallic to charge ordered insulating while increasing *z* [15], with a coexisting phases scenario at intermediate doping. The ground state phase diagram in the *x-z* space was also presented, showing the existence of FM, CO and PS states[15]. Evidences of the phase separated character in several compounds of the series were also found using Raman spectroscopy [16-18]. The phase separated state induced by the chemical substitution with Nd, Gd and Y was studied by L. Sudheendra et al. [19]; the authors showed that disorder, related with variance of the mean radius of the A – site ion, favors the AFM phase. Regarding the dynamics of the phase separated state, F. Rivadulla et al. [20] showed that the low temperature state constitutes a self generated ensemble of magnetic clusters that behaves like a spin glass in $(La_{0.25}Nd_{0.75})_{0.7}Ca_{0.3}MnO_3$.

In this work we present a detailed study based on the electrical and magnetic properties of the $La_{5/8-y}Nd_yCa_{3/8}MnO_3$ series focusing specially on the characterization of the physical response of the system within the phase separated regime. As it was shown, [11] in the PS regime strong dynamical effects are responsible for the blurring of the real equilibrium state of the system. This dynamical regime is characterized by the strong difference between the magnetization measurements in zero field cooling (ZFC) and field cooled (FC) modes and in the existence of blocking temperatures, resembling the behaviour of spin glasses.[11] We have used our data to construct the electric and magnetic phase diagram of the system, including the existence of frozen and dynamic regimes suggested by these measurements. We have paid special attention in a detailed description of the low temperature phase separated state corresponding to intermediate doping.

## 2. Experimental

Bulk polycrystalline samples of $La_{5/8-y}Nd_yCa_{3/8}MnO_3$ (LNCMO(y)) were synthesized following the liquid mix method (citrate polymerization). The obtained powder was pressed into bars with the dimension of $5 \times 1 \times 1$ mm$^3$. The X-ray diffraction pattern corresponds to a single phase material, with no spurious lines observed. DC magnetization measurements were performed in a Quantum Design PPMS system. Resistivity was measured in a He closed cycle criogenerator using the standard four probe method.

## 3. Results and Discussion

Figure 1 shows the DC magnetization ($M$) and resistivity ($\rho$) as a function of temperature ($T$) for LNCMO($y$) samples with $y$ = 0.1, 0.2, 0.3, 0.4, 0.5 and 0.625. $M(T)$ was measured in the field cooled cooling (FCC) mode using an applied magnetic field of 0.2 T.

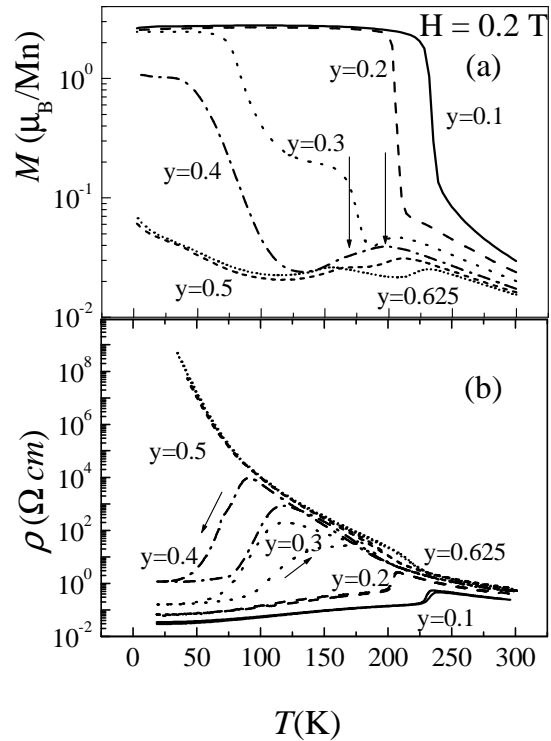

**Figure 1: (a) $M$ (FC with H = 0.2 T) and (b) $\rho$ vs $T$ for LNCMO(y) with y=0.1, 0.2, 0.3, 0.4, 0.5 and 0.625. The arrows indicate two magnetic transitions for the y=0.4 sample.**

In the low doping limit ($y$ = 0.1 and 0.2), the samples exhibit paramagnetic and insulating (PI) behaviour at high temperatures. Both samples display a typical PI to FM – metallic transition on cooling below $T_C$, resembling the characteristics of the $La_{5/8}Ca_{3/8}MnO_3$ parent compound [21]. The increment of the Nd content produces a reduction of $T_C$ and an increase of the low temperature resistivity, indicating that doping destabilizes the FM state. A small irreversibility is present on the $\rho$ vs. $T$ dependence of LNCMO(0.2) around 100 K (not apparent due to the scale of the graph). Moreover, a complex behaviour is observed both on $M$ and $\rho$ upon further doping increase. Samples with $y = 0.3$ and 0.4 became less FM and much more resistive at low temperatures and a clear hysteresis loop is observed in the $\rho$ vs. $T$ dependence. On cooling, LNCMO(0.3) shows a reduction of $M$ that is correlated with a maximum of the logarithmic derivative of the resistivity, $\partial \ln \rho / \partial (1/T)$ which indicates the arising of the CO state at $T_{co} \sim$ 205 K. The onset of FM order occurs at lower temperatures, ~180 K, but the transition is not completed as can be seen by the plateau in $M$ between 100 and 175 K on cooling. This data is similar to that obtained for the $La_{5/8-y}Pr_yCa_{3/8}MnO_3$ system [3,5,6] in which it has been shown to be related to PS, so this suggests that the LNCMO(0.3) compound is phase separated in FM and CO regions between 175 and 90 K. Below $T_C \approx 90$ K magnetization reaches almost the same value than the corresponding to the pure FM compounds ($0 < y < 0.2$) at low temperature, indicating a majority FM state in this range. This transition coincides with the change from insulator to metallic behaviour indicated by a peak in the $\rho$ vs. $T$ dependence at $T_p$.

The LNCMO(0.4) sample shows a similar behaviour, but with lower values of $M$ and a higher resistivity at low temperatures. As a distinctive feature, this compound presents two reductions of the magnetization on cooling indicated by vertical arrows in figure 1(a), one that coincides with $T_{co}$ and the other at $T \sim 175$ K, not correlated with any electrical transport feature. This peculiarity was attributed on the $Pr_{1-x}Ca_xMnO_3$ system to the appearance of an AFM state [22]. This sample presents an increase on the magnetization on cooling below ~ 100 K but its low temperature value is lower than the corresponding to a pure FM system, suggesting coexistence of FM and non FM phases. The resistivity this sample also displays a maximum at $T_p$, which marks the transition from insulating to metallic behaviour.

The $y = 0.5$ and 0.625 samples show an insulating behaviour in the whole temperature range studied (see figure 1(b)). They are paramagnetic at high temperature and undergo a transition to a CO state at 210 K and 230 K respectively. On further cooling, both samples show a kink on the $M$ vs $T$ data which is presumably due to the AFM ordering within the CO phase [22]. The comparatively low magnetization value at low temperature for both compounds is an indication of a majority AFM-CO phase. On cooling below 100 K we can see that the magnetization of both samples shows a small upturn (figure 1(a)). This behaviour is similar to that corresponding to the magnetization curves obtained by Dupont et al. for $Nd_{1-x}Ca_xMnO_3$ manganites [23]. By analyzing electron spin resonance and neutron diffraction data, the authors claim that this feature can be related to the paramagnetic moment of the Nd ions within the AFM state of the Mn system.

On figure 2 we show resistivity as a function of applied field measurements, $\rho$ vs. $H$, at fixed T ~ 90 K for LNCMO(0.4). Negative magnetoresistance (MR) is observed for all magnetic fields, with two well differentiated regimes: below 0.4 T we have the "classical" MR effect due to the enhanced itineracy of the electrons in the FM state [1] while for larger fields, a change in the slope indicates a reduction of $\rho$ related to the field induced increase of the relative FM fraction, an effect that can only happen in a phase separated system. It has been shown that the later effect can produce a change in the resistivity and magnetization which persists after the field is turned off [24,25]. Remarkably, this nonvolatile cumulative effect is observed while cycling and increasing the maximum field. The preceding result suggests us that LNCMO samples with $y$ near 0.4 exhibit PS around 90 K. Magnetic field was cycled to be sure that the nonvolatile reduction of the resistivity while turning off the field is not due to FM remanescence but only related with the maximum applied magnetic field.

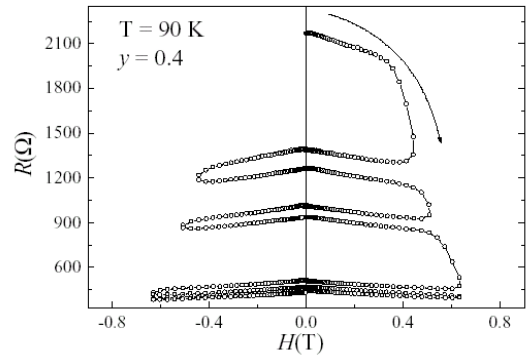

**Figure 2: Resistivity as a function of the magnetic field for LNCMO(0.4) at ~90 K.**

Detailed magnetic measurements provide further insight on the nature of the phase separated state. The $M$ vs $H$ dependence for LNCMO(0.3) below $T_C$ after ZFC from room temperature is shown in figure 3. At first sight, the data show the typical shape corresponding to a soft and homogeneous FM sample. However, small traces of hysteresis signal the presence of inhomogeneities, even at temperatures well below 90 K, i.e. a phase separated state with a majority FM fraction (see figure 1(a)). Similar measurements on LNCMO(0.4) depicted in figure 4, show the development of the phase separated state on cooling. At 106 K we see no signature of the FM phase according to the low susceptibility near $H = 0$, while saturation can be reached just above 3 T. The existence of the FM phase at zero field is evident at $T = 70$ K and a clear hysteresis is displayed. The global behaviour at 2.5 K is qualitatively similar, with an increasing FM fraction at zero field and the irreversible metamagnetic transition due to the FM / AFM coexistence. An interesting point is that the transition at 2.5 K is very abrupt. Similar transitions were already observed in the $La_{5/8-y}Pr_yCa_{3/8}MnO_3$ (y=0.4) compound [26], in which the application of a large enough magnetic field the can transform the low temperature zero field cooled mostly CO state, into a nearly homogeneous FM state in an abrupt step-like metamagnetic transition, which remains stable even after the field is removed. The basic condition for the occurrence of the abrupt transition is that the system must reach the low temperature regime in a strongly blocked state; the interplay between the field induced growth of the FM phase against the CO one and the heat released by this process seems to be the key features to explain its occurrence [26].

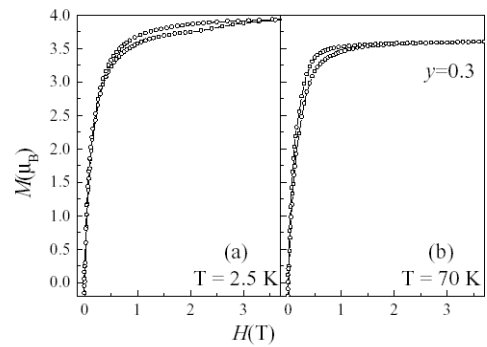

**Figure 3:** *M* vs *H* for LNCMO(0.3) at (a) 2.5 K and (b) 70 K.

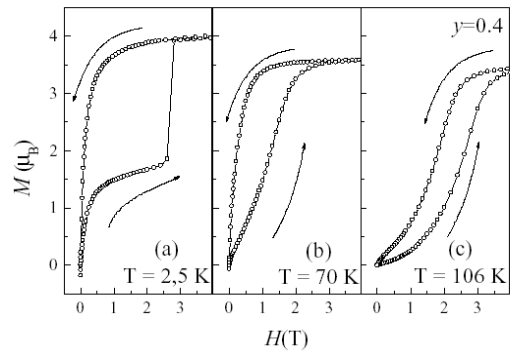

**Figure 4:** *M* vs *H* for LNCMO(0.4) at (a) 2.5 K, (b) 70 K and (c) 106 K. The abrupt metamagnetic transition is observed at 2.5 K for $H = 2.7$ T.

Figure 5 shows the *M* vs *H* dependence at 2.5 K for LNCMO(0.5) (figure 5(a)) and LNCMO(0.625) (figure 5(b)), showing no evidence of the FM phase at null *H*. This last data, along with the previous presented *M* vs *T* dependence for these samples, in which no transition to a FM nor metallic state was found, show that the two compounds present an homogeneous AFM-CO phase for low $H$ values. An abrupt metamagnetic transition similar to that shown for LNCMO(0.4) is observed at around $H$ = 5.5 T for LNCMO(0.5) and $H$ = 7.9 T for LNCMO(0.625) indicating the existence of a blocked or frozen state at low temperature for the two compounds.

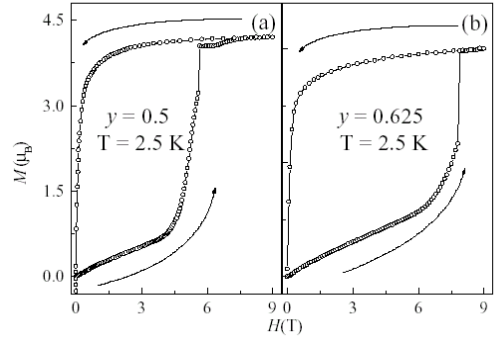

**Figure 5: $M$ vs $H$ for (a) LNCMO(0.5) and (b) LNCMO(0.625) at 2.5 K. Abrupt metamagnetic transitions are observed at $H$ = 5.5 T and 7.9 T respectively.**

Previous studies have shown evidences of the tendency of the PS state in manganites towards a characteristic dynamic behaviour. On general grounds, when two phases coexist, competition of interactions that tend to stabilize one or the other is expected, which can eventually lead to frustration and thereby to glassy features. In addition, frustration can arise due to the structural differences between the coexisting phases.

Figure 6 shows the $M(T)$ dependence, using $H$ = 0.2 T in FC and ZFC modes for samples with $y \geq 0.3$. The difference between ZFC and FC data suggest that the ZFC state of the samples at low temperature is metastable, i.e. frozen. When temperature is increased in the presence of a magnetic field, the system unblocks and the FM phase grows against the AFM-CO one. We identify the temperature where the ZFC curve has a maximum derivative with a blocking temperature ($T_B$) which depends on the applied magnetic field. A similar behaviour was observed recently on $La_{5/8-y}Pr_yCa_{3/8}MnO_3$ with y = 0.4 [11] and y = 0.375 [27]. In ref. [11], it has been shown that the ZFC/FC difference and a characteristic dynamical behaviour can be well accounted for by proposing a temporal evolution of the relative fraction of the coexisting phases through a hierarchy of energy barriers, which only depend on the actual value of these fractions. This dynamic behaviour is attributed to phase competition rather than solely to magnetic interaction frustration as happens in conventional spin glasses. The abrupt metamagnetic transition observed for y = 0.4, 0.5 and 0.625 is also a signature that the systems are blocked at low temperature.

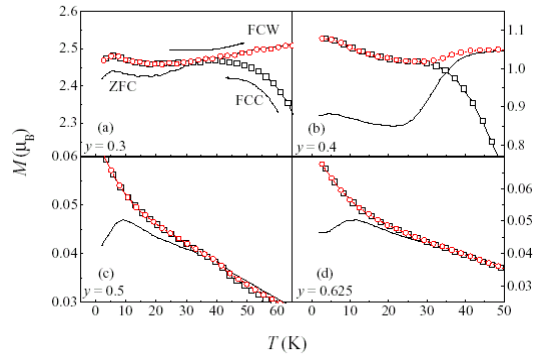

**Figure 6: M vs T for y = 0.3, 0.4, 0.5 and 0.625 samples in the FC and ZFC modes.**

In figure 7 (a) and (b) we show the dependence of $T_B$ for $H$ = 0.2 T and the difference $M_{FC}$ - $M_{ZFC}$ = $\Delta M_{FC\text{-}ZFC}$ measured at 2 K, respectively, as a function of Nd doping. Both variables exhibit a peak for $y$ = 0.4. The blocking temperature can be regarded as a measure of the thermal energy needed for the system to overcome the energy barriers that separate the FM and AFM states; a larger $T_B$ means higher barriers and thus a larger tendency to be blocked or "frozen". On the other hand, $\Delta M_{FC\text{-}ZFC}$ measured at low temperatures is expected to be large when the system has a considerable volume fraction in the blocked regime. An interesting point is that a large value of $T_B$ (large energy barriers) is correlated with a large $\Delta M_{FC\text{-}ZFC}$ (large difference between the volume fractions in the blocked and non blocked regimes). The maximum of both quantities near y = 0.4 is the signature of the enhanced blocked behaviour below $T_B$. It is worth to note that the existence of blocking temperatures is one of the most characteristic and intriguing features of phase separated manganites [11,27,28].

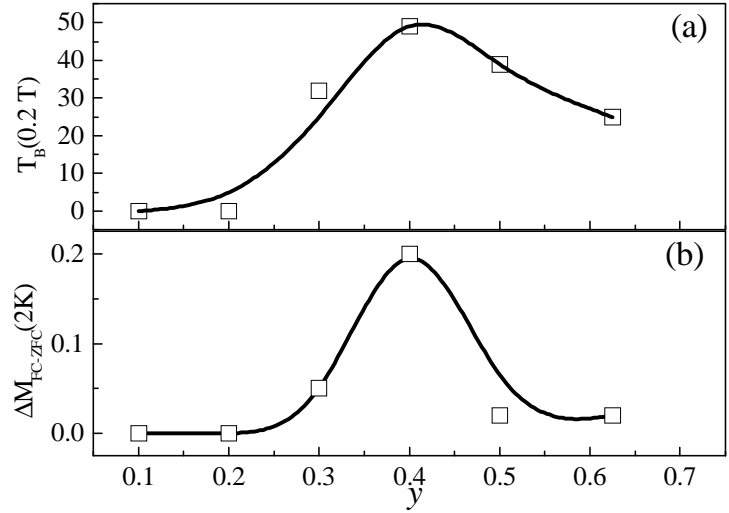

**Figure 7: (a) Blocking temperature ($T_B(0.2\ T)$) and (b) the difference of $M_{FC}(2K)$ and $M_{ZFC}(2K)$, as a function of doping.**

The overall presented results can be summarized in the phase diagram displayed on figure 8. The filled symbols correspond to data obtained from *M(T)* results while the empty symbols were obtained from *ρ(T)*. The magnetic transition temperatures obtained are in well agreement with those presented by Rao et al. [14] and Moritomo [15], displaying the same overall tendency on doping, although in our case they are slightly smaller. Regarding Rao's paper, the author has also observed a temperature corresponding to the change from insulator to metallic behaviour ($T_p$). Our results for $T_p$ are shown in empty circles in fig. 8. All samples are PI at high temperature. In the low Nd doping limit, the system undergoes a transition to the FM metallic state on cooling below 230 K. On doping, the system becames less FM as can be inferred from the reduction of $T_C$, and the CO state tendency is enhanced. These effects are due to the increase of the distortions of the Mn – O – Mn bond angles while reducing $<r_A>$. At the other side of the diagram, the "pure Nd" compound first becomes charge ordered at 225 K and on further cooling it orders antiferromagnetically ($T_N \sim 150$ K).

In the intermediate doping region, limited by dashed lines, low temperature data showed evidences of the phase separated character of the samples. Within the PS regime, the change from insulator to metallic behaviour at $T_p$ mark the limit between the temperature ranges in which the FM fraction may form percolative paths (percolative PS or PPS) or not (non percolative PS or NPPS) [14]. The existence of a blocked regime, enhanced around $y = 0.4$, shows that the phase coexistence at low temperatures is strongly affected by dynamical features of the phase separated state, which allows the existence of frozen metastable states (frozen PS).

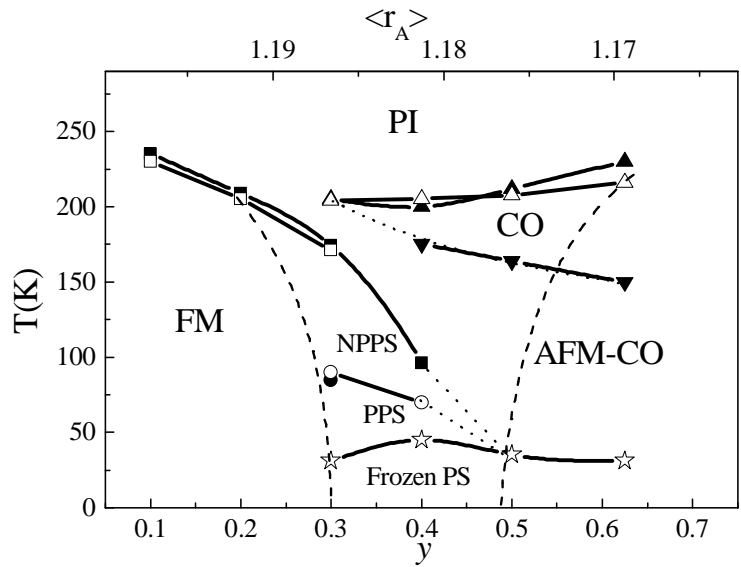

**Figure 8: Electric and magnetic phase diagram of the LNCMO(y) system. Data of filled symbols are taken from *M(T)* while empty ones are extracted from *ρ(T)*. The dashed lines mark the separation between homogeneous and the PS state. Percolative PS (PPS) and non percolative PS (NPPS) stand for a state in which the FM metallic fraction form percolative paths through the sample or not, respectively.**

## 4. Conclusions

In conclusion, we performed magnetization and transport measurements on the $La_{5/8-y}Nd_yCa_{3/8}MnO_3$ system in order to study the influence of the A – site mean ionic radius on the physical properties. The reduction of $<r_A>$ inhibits the FM state, because it increases the number of distortions of the Mn – O – Mn bonds, and thus favors charge localization. The existence of a critical radius $<r_A> \sim 1.18$ Å where phase separation appears is evident, as previously observed in ref. [19]. Our results show that the system is a paramagnetic insulator at high temperatures, while on cooling, the ground state is FM metallic for low doping and AFM CO for high doping. For intermediate doping, there is a zone where the most stable state is phase separated. The change on the electrical transport behaviour from insulator to metallic, show the temperature below which the FM phase form percolative paths through the sample.

The dashed lines on figure 8 mark the separation between homogeneous states and the phase separated state. According to recent theoretical developments, the origin of this window is closely related to the existence of disorder [12,13]. In this case, disorder comes from the random distribution of the La and Nd ions on the A – sites. The presence of smaller Nd ions causes distortions of the Mn – O – Mn bond angle which will be also disordered. These distortions can produce long range deformations decaying as $1/r^3$, that trigger the clustering tendencies leading to the formation of the phase separated state [29]. A particular dynamical behaviour was also observed inside this window characterized by the difference between the ZFC and FC curves, the existence of blocking temperatures and abrupt metamagnetic transitions. According to the particular phenomenology of phase separated manganites, the competition between phases appear to be the cause of these behaviour. This dynamic phase separated state was also theoretically proposed and the presence of disorder is believed to be the key for its existence [12,13].

**Acknowledgements**

The authors thank M. Quintero and L. Granja for fruitful discussions. P. Levy is a member of CIC CONICET. This work was partially supported by ANPCyT (PICT03-13517) and Fundación Antorchas (Argentina); and CAPES, FAPERJ, and CNPq (Brazil).